\documentstyle[aps,prl,multicol,epsfig,amssymb]{revtex}

\title{Stability of inverse bicontinuous cubic phases in lipid-water mixtures}
\author{U. S. Schwarz${}^{1}$ and G. Gompper${}^{2}$}
\address{${}^{1}$Department of Materials and Interfaces,
Weizmann Institute of Science,
Rehovot 76100, Israel \\
${}^{2}$Institut f\"ur Festk\"orperforschung,
Forschungszentrum J\"ulich,
52425 J\"ulich, Germany}   

\begin{document}

\maketitle

\begin{abstract}
  We investigate the stability of seven inverse bicontinuous cubic
  phases ($G$, $D$, $P$, $C(P)$, $S$, $I-WP$, $F-RD$) in lipid-water
  mixtures based on a curvature model of membranes. Lipid monolayers are
  described by parallel surfaces to triply periodic minimal surfaces.
  The phase behavior is determined by the distribution of the Gaussian
  curvature on the minimal surface and the porosity of each structure.
  Only $G$, $D$ and $P$ are found to be stable, and to coexist along 
  a triple line.  The calculated phase diagram agrees very well with 
  experimental results for 2:1 lauric acid/DLPC.
\end{abstract}
\pacs{PACS numbers: 68.10.-m, 61.30.Cz, 87.16.Dg}


\begin{multicols}{2}
\narrowtext                                                       

Most of the many mesomorphic phases formed by lipid-water mixtures are
of the inverse type, with the self-assembled lipid monolayers 
curving towards the aqueous regions \cite{exp}. In inverse
bicontinuous cubic phases, a \emph{single} lipid bilayer extends
throughout the whole sample, dividing it into two percolating water
labyrinths. Until now, the structures $G$, $D$ and $P$ have been identified.
The only known lipid-water system in which $G$, $D$ and $P$
coexist is 2:1 lauric acid/DLPC and water \cite{templer}. The property
of cubic lipid bilayer phases to divide space into interwoven polar
and apolar compartments is utilized for biological function, e.g.\ in
mitochondria and the endoplasmic reticulum \cite{bio}. Recently, they
have also been used as artificial matrices which enable membrane
proteins like bacteriorhodopsin to crystallize in a three-dimensional
array \cite{a:peba97}. 

\begin{figure}[h]
\begin{center}
\epsfig{figure=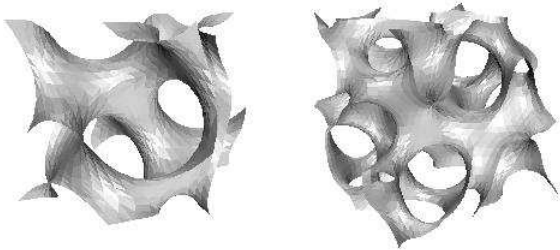,width=0.9\columnwidth} 
\end{center}
\caption{One of the two monolayers in the $G$ (left) and $S$ (right)
  structures, for hydrocarbon volume fraction $v=0.5$.
  Both inverse bicontinuous cubic phases have spacegroup $Ia\bar3d$.
  However, only $G$ is found to be stable both in experiments and
  theory.}
\label{FigurePictures}
\end{figure}

It was shown by Luzzati and coworkers \cite{a:mari88} that the
mid-surfaces of the lipid bilayers are very close to cubic {\em minimal
surfaces}, which have zero mean curvature everywhere.  
These surfaces occur in lipid-water systems due to the local
symmetry of the lipid bilayer, which implies that the surface should
curve to both sides in the same way.  However, it is well known
\cite{ms} that many more cubic minimal surfaces exist than the
structures $G$, $D$ and $P$ identified in lipid-water mixtures. What
might be the reason that these other phases have not been observed so
far?  Helfrich and Rennschuh \cite{a:helf90} have argued, based on the
curvature model for fluid membranes \cite{curvature}, that structures
with a narrow distribution of Gaussian curvature over the minimal
mid-surface should be most favorable.  However, the relevant data was
known to them only for $G$, $D$ and $P$, which are degenerate in this
respect due to the existence of a Bonnet transformation. Recently, we
obtained numerical representations for a large number of cubic minimal
surfaces in the framework of a simple Ginzburg-Landau model
\cite{uss:schw99}.  In this Letter we use this data to investigate
seven inverse bicontinuous cubic phases ($G$, $D$, $P$, $C(P)$, $S$,
$I-WP$, $F-RD$). For an illustration of the $G$ and $S$ surfaces, see
Fig.~\ref{FigurePictures}. We find that the width of the different
distributions of Gaussian curvature is indeed smallest for $G$, $D$
and $P$ and larger for all other structures considered. This proves
for the first time why only $G$, $D$ and $P$ should be observed
experimentally.

Our detailed investigation of the stability of bicontinuous cubic
phases shows that the existence of the Bonnet transformation implies
that with increasing water concentration, $G$, $D$ and $P$ coexist
along a triple line. We show that this sequence is determined by an
universal geometrical quantity, the topology index, and that higher
order terms in the curvature energy and thermal membrane undulations do 
not lift this degeneracy. Furthermore, we calculate phase diagrams as
function of concentration and temperature, which are in good agreement
with an intermediate temperature range of the experimental phase
diagram for 2:1 lauric acid/DLPC and water \cite{templer}.

The main contributions to the free energy of an inverse bicontinuous
cubic phase are the curvature energy of the lipid monolayers and the
stretching energy of the hydrocarbon chains \cite{parallel}. The
curvature energy is described by the {\em Canham-Helfrich Hamiltonian} 
\cite{curvature}
\begin{equation} \label{HelfrichHamiltonian}
{\cal H} = \int dA\ \left\{ 2 \kappa\ (H - c_0)^2 + \bar \kappa\ K \right\}\ ,
\end{equation}                                                            
where $H$ and $K$ are mean and Gaussian curvature, respectively, and
the integral runs over the neutral surface of the monolayer. The model
parameters are spontaneous curvature $c_0$, bending rigidity $\kappa$
and saddle-splay modulus $\bar\kappa$. In the following, positive mean
curvature corresponds to curvature towards the aqueous regions.  The
stretching energy can be taken to be harmonic about the average chain
extension $l$. In order to attain optimal free energy, the neutral
surfaces of the two monolayers should realize at the same time
constant mean curvature $c_0$ and constant distance $l$ to the minimal
mid-surface.  Since this is not possible geometrically, the 
system will always be frustrated. The relative distribution of the
frustration on bending and stretching energies depends on $l$. In a
simple polymeric model (which certainly overestimates the chain 
compressibility of lipid monolayers), bending rigidity and stretching 
modulus scale
as $l^3$ and $l^{-1}$, respectively \cite{a:safr99}, and the ratio of
the two energies scales as $(l/a)^2$, where $a$ is the lattice
constant. For low water content, where 
$a\simeq 2l$, stretching is important and the
bicontinuous cubic phases are suppressed by the hexagonal phase. 
In this Letter we consider the opposite limit, 
where the stretching energy dominates and the layer thickness is
nearly constant. We
model the two monolayers as parallel surfaces at distances $\pm l$
away from the minimal mid-surface.  Their geometrical properties are
thus given by
\begin{equation} \label{ParallelSurfaces}
dA^l = dA\ (1 + K l^2)\ , 
H^l  = \frac{- K l}{1 + K l^2}\ , 
K^l  = \frac{K}{1 + K l^2}\ .
\end{equation} 
With these relations, the curvature energy of the lipid bilayer can be
expressed in terms of the geometrical properties of the minimal
mid-surface. 

Our analysis shows that each (inverse) bicontinuous cubic phase is
characterized by two dimensionless numbers, $\Delta$ and $\Gamma$,
which are sufficient to describe its stability and phase behavior.
$\Delta$ is the width of the distribution of Gaussian curvature over
the surface and will be discussed in more detail below. The topology
index $\Gamma$ can be motivated as follows.  For a two-dimensional
surface embedded in three-dimensional space, there exist three
motion-invariant and additive functionals of the shape: surface area
$A$, Euler characteristic $\chi$ and integrated mean curvature.  For
triply periodic minimal surfaces, the integrated mean curvature
vanishes. The values for $A$ and $\chi$ depend on lattice constant and
choice of unit cell, but one ratio can be constructed which does not. We
define this universal quantitiy to be the topology index $\Gamma =
\left( A^3 / 2 \pi |\chi| V^2 \right)^{1/2}$, where $V$ is the volume
of the unit cell. It describes the porosity of a structure (the larger
its value, the smaller is the connectivity) and has been investigated
extensively by Hyde and coworkers \cite{hyde}.  The gyroid $G$ is the
least porous of the known triply periodic minimal surfaces since it
divides space into labyrinths with 3-vertices only.  Then comes $S$
which has both 3- and 4-vertices, and $D$, which has 4-vertices only.
All other structures involve higher-coordinated vertices.  In Table~I
we collect values for $\Gamma$ for all cubic minimal surfaces
considered in this Letter. The values for $G$, $D$, $P$ and $I-WP$ are
exact since exact (Weierstrass) representations are known.  $G$, $D$
and $P$ are related by a Bonnet transformation, which changes the
geometrical properties globally (i.e.\ shape in embedding space,
topology, space group and lattice constant) but not locally 
(i.e.\ metric and curvatures). The values of $\Gamma$ for $C(P)$, $S$ and
$F-RD$ given in Table~I follow from our numerical representations as
isosurfaces of scalar fields $\Phi(r)$ which we obtained recently from
a simple Ginzburg-Landau model for ternary amphiphilic systems
\cite{uss:schw99}; due to numerical limitations, they are not too
accurate but can be expected to appear at the correct position in the
overall sequence.

For a given cubic minimal surface with surface area $A^*$ and Euler
characteristic $\chi$ per conventional unit cell, the hydrocarbon
volume fraction $v$ of the corresponding inverse bicontinuous cubic
phase is given by
\begin{equation} \label{HydrocarbonVolume}
v = \frac{1}{a^3} \int_{- l}^{l} d{l'} \int dA^{l'}
= 2 A^* \left( \frac{l}{a} \right) 
       + \frac{4 \pi}{3} \chi \left( \frac{l}{a} \right)^3\ .
\end{equation}   
It turns out that for all seven structures considered the lattice
constant in units of chain length is very well approximated by $a / l
= 2 A^* / v$ over the whole range of $v$ (for $v \gtrsim 0.8$, the
bicontinuous structures begin to self-intersect). The average mean
curvature of the monolayers as a function of $v$ is
\begin{equation} \label{MeanCurvature}
\langle H^l \rangle_l l = \frac{\int dA^l\ H^l l}{\int dA^l}
= \frac{(v / \Gamma)^2}{4 - (v / \Gamma)^2}\ .
\end{equation} 
In the following, spontaneous curvature $c_0$ is measured in units of
$1/l$. Eq.~(\ref{MeanCurvature}) implies that a given
spontaneous curvature is attained at the hydrocarbon volume fraction
$v = ( 4 c_0 / (1 + c_0) )^{1/2}\ \Gamma$. Hence if all structures
considered were stable, for increasing water concentration we would
expect them to appear in a sequence of decreasing topology
index, i.e.\ in the sequence $G$ - $S$ - $D$ - $I-WP$ - $P$ etc.

The stability of the different phases is determined by the free energy
per unit volume. In the absence of thermal fluctuations, the free
energy density can be written in dimensionless units as
\begin{eqnarray} \label{FreeEnergyDensityBicont}
f_B & = & v\ \Big\{ \int \frac{dA^*}{A^*}
\left( 1 -  \Xi(K^*) \left(\frac{v}{\Gamma} \right)^2 \right)^{-1} \\
& & \times 
\left( 1 - \frac{1+c_0}{c_0} \Xi(K^*) \left(\frac{v}{\Gamma} \right)^2 \right)^2
+ \frac{r}{4 c_0^2} \left( \frac{v}{\Gamma} \right)^2 \Big\} \nonumber
\end{eqnarray}             
where the integration extends over the minimal mid-surface in a
conventional unit cell. Here we defined $r = - \bar \kappa / 2 \kappa$
and $\Xi(K^*) = K^* A^* / 8 \pi \chi$. The curvature model
(\ref{HelfrichHamiltonian}) is well-defined for $0 \le r \le 1$.
$\Xi(K^*)$ varies over the surface, but can be shown to be invariant
under a Bonnet transformation. Higher-order terms in the curvature
energy of the monolayers would result in expressions involving higher
powers of the combination $\Xi(K^*) (v / \Gamma)^2$.  The free energy
density of the lamellar phase is $f_{L_{\alpha}} = v$.

Eq.~(\ref{FreeEnergyDensityBicont}) can also be expressed in the form
of an effective curvature energy for the lipid bilayer (like
Eq.~(\ref{HelfrichHamiltonian}) for the monolayers).  From an
expansion for small chain length $l$, one obtains an infinite series
in powers of Gaussian curvature $K$. In particular, the effective
value for the saddle-splay modulus of the lipid bilayer is found to be 
$\bar\kappa_{bi} = 2 (\bar \kappa + 4 c_0 \kappa)$.
Since $\bar\kappa_{bi}$ increases linearly with the spontaneous curvature 
of the monolayers, the cubic bicontinuous phases are favored for 
$c_0 > r/2$, even though $\bar\kappa$ of the monolayers is assumed to
be negative.  

The first (bending) term in Eq.~(\ref{FreeEnergyDensityBicont}) measures
the deviation of the monolayers' mean curvature from the spontaneous
curvature.  It follows from Eq.~(\ref{ParallelSurfaces}) that a narrow
distribution of Gaussian curvature $K$ over the minimal mid-surface
translates into a narrow distribution of mean curvature $H^l$ over the
monolayer, which is favorable in terms of the bending term, as pointed
out by Helfrich and Rennschuh \cite{a:helf90}. For a more detailed
analysis, we note that the bending term is an area average of a
complicated function of its Gaussian curvature $K$, which is
distributed non-uniformly over the surface.  We proceed by first
measuring the distribution $f(K)$ of Gaussian curvature over the
minimal surface.  Even in the case of a known Weierstrass
representation, the function $f(K)$ is analytically not tractable and
has to be determined numerically in the form of a histogram
$\{K_i,f_i\}$. Its knowledge allows to replace the area integral in
Eq.~(\ref{FreeEnergyDensityBicont}) by a (one-dimensional) integral
over Gaussian curvature.  We obtained $\{K_i,f_i\}$ for $G$, $D$, $P$
and $I-WP$ from their Weierstrass representations and for $C(P)$, $S$
and $F-RD$ from our numerical representations. A short description of
the procedure in both cases and plots of the histograms obtained are
given in Ref.~\cite{uss:schw99}.  In order to quantify the width of
the different distributions, we calculate $\Delta = \langle ( K -
\langle K \rangle )^2 \rangle / \langle K \rangle^2$ where the average
is over the surface and normalized.  Like the topology index $\Gamma$,
$\Delta$ is independent of scaling and choice of unit cell. Moreover,
it can be shown that it is also invariant under Bonnet
transformations. The values for $\Delta$ for the seven structures
considered are given in Tab.~I. We find that it is the smallest for
$G$, $D$ and $P$ and higher for all other structures.  Thus we
conclude that although for given $c_0$ the different structures have
different optimal values for $v$ according to their topology index,
only $G$, $D$ and $P$ will have small frustration due to the
relatively small variation of mean curvature over their monolayers.

Since the quantity $\Xi(K^*)$ in Eq.~(\ref{FreeEnergyDensityBicont})
is invariant under a Bonnet transformation, the free energy densities
for $G$, $D$ and $P$ obey a scaling form $v g((v/\Gamma)^2)$, with a
universal function $g(x)$ which can be read off from
Eq.~(\ref{FreeEnergyDensityBicont}).  For small $r$, this function has
$g(0) = 1$, a minimum at $x \simeq 4 c_0 / (1+c_0)$, and diverges at
$x \simeq 4$. An analytical Maxwell construction for this scaling
function shows that the Bonnet-related structures form a triple line,
such that the coexisting values of $v$ scale with $\Gamma$. Moreover
the triple line includes an excess water phase at $v = 0$. The
coexistence with an excess phase is known as \emph{emulsification
  failure} from surfactant systems \cite{a:safr99}: if the swelling
with water reaches the point where the structure adapts best to the
given spontaneous curvature, any additional water is just expelled
into an excess phase.

In addition to the bending energy, {\it thermal fluctuations} have to be
taken into account for the calculation of phase diagrams. In the
case of the lamellar phase, the membrane undulations give rise to a 
steric repulsion \cite{a:helf78}, which in our case reads
\begin{equation}
f_{steric} = \frac{c_\infty}{32 c_0^2} \ 
\left(\frac{k_BT} {\kappa}\right)^2 \ \frac{v^3}{(1 - v)^2} 
\end{equation}
where $c_\infty = 0.106$ \cite{a:gomp89b}.  In the bicontinuous cubic
phases, the fluctuations of the mid-plane lead to a renormalization of
$\kappa$ and $\bar\kappa$ \cite{a:nels89}. However, since $\kappa$
multiplies the average mean curvature squared, which vanishes for the
mid-plane, only the renormalization of $\bar\kappa$ at length scale
$\ell$ has to be taken into account.  We identify the typical length
scale of a cubic structure with $\langle K \rangle^{-1/2}$. This
implies $\ell/l = 2 \Gamma / v$, so that $r$ in
Eq.~(\ref{FreeEnergyDensityBicont}) gets renormalized, with
\begin{equation}
r_R = \left[ r - \frac{5}{12\pi} \frac{k_BT}{\kappa} 
 \ln\left(\frac{2 \Gamma}{v} \right) \right]\ .
\end{equation}
Therefore the scaling form of the free energy remains unchanged both
by higher order terms in the curvature energy and by thermal
undulations, although the scaling function itself changes.
 
\begin{figure}
\begin{center}
\epsfig{figure=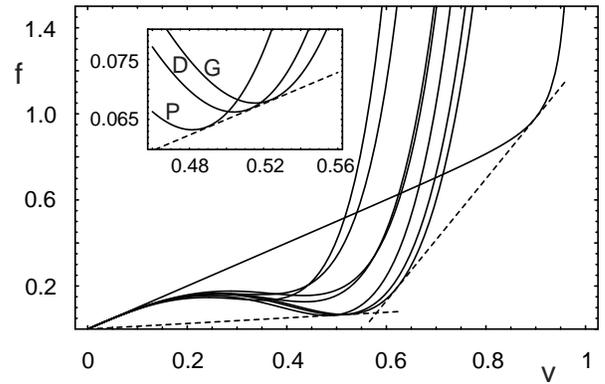,width=0.9\columnwidth}
\end{center}
\caption{Free energy densities as a function of hydrocarbon volume 
  fraction $v$ for $c_0 = 1/6$, $r=0$ and $\kappa/k_BT=10$.  The
  phases can be identified from the sequence of curves at the top,
  which corresponds to $F-RD$, $C(P)$, $I-WP$, $S$, $P$, $D$, $G$ and
  $L$ from left to right.  Only $G$, $D$ and $P$ are stable; they form
  a triple line (see inset) which includes a coexistence with excess 
  water (lower dashed line). The second
  dashed line is a Maxwell construction between $G$ and the lamellar
  phase.}
\label{FigureFreeEnergy}
\end{figure}

\begin{figure}
\begin{center}
\epsfig{figure=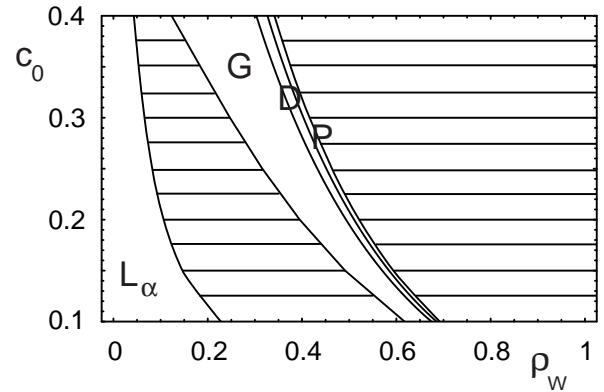,width=0.9\columnwidth} 
\end{center}
\caption{Numerically calculated phase diagram for $r = 0.1$ and 
  $\kappa/k_BT=10$ as a function of water volume fraction $\rho_W=1-v$
  and spontaneous curvature $c_0$. The hatched regions indicate two-phase 
  coexistences.}
\label{FigurePhaseDiagram}
\end{figure}

In Fig.~\ref{FigureFreeEnergy}, we plot the free energy densities for
$c_0 = 1/6$, $r = 0$ and $\kappa/k_BT = 10$.  As predicted from our
general results above, we find that $G$, $D$ and $P$ are indeed the
only stable cubic phases and that they become stable along a triple
line, which is the lower straight line.  The same line also marks the
Maxwell construction for the emulsification failure, the coexistence
of P with an excess water phase. These curves
for different values of $c_0$ allow the construction of phase diagrams
for given $r$ and $\kappa/k_BT$ as shown in
Fig.~\ref{FigurePhaseDiagram}.  Note that without any further
interactions, $D$ is only stable along a line in our calculation.  We
therefore do not show two-phase coexistence regions between the three
bicontinuous cubic phases in Fig.~\ref{FigurePhaseDiagram}, but only
the intersections of the free-energy curves.  With increasing $c_0$,
the regions of stability shift to higher values of $v$ as predicted by
Eq.~(\ref{MeanCurvature}).  

The existence of the triple line implies that subtle differences in the 
free energies, which arise from a more detailed treatment of stretching 
contributions, of thermal fluctuations, or of van der Waals or 
electrostatic interactions, can be expected to destroy the degeneracy 
between $G$, $D$ and $P$. This explains qualitatively why all three 
cubic phases are rarely observed in the same system.

We can now compare our results with the experimental phase diagram for
2:1 lauric acid/DLPC and water, the only known experimental system in
which $G$, $D$ and $P$ coexist. Since the head group region is much
smaller than the tail region of the lipid, we can identify the
hydrocarbon volume fraction $v$ with the lipid concentration.
Further, a linear relation between $c_0$ and $T$ can be assumed, since
it is well established in non-ionic surfactant systems.  In both phase
diagrams, the sequence $L$ - $G$ - $D$ - $P$ - $W+P$ is observed with
increasing water concentration $\rho_W \approx 1 - v$.  For $r = 0.1$,
the temperature range from $T = 35^0 C$ to $T = 50^0 C$ in the phase
diagram of 2:1 lauric acid/DLPC and water corresponds roughly to the
range $0.18$ to $0.26$ in spontaneous curvature $c_0$.
Experimentally, the hexagonal phase dominates at higher temperatures
and large hydrocarbon volume fraction, and fluidity is lost at lower
temperatures (with the main transition at $30^o C$).  An extrapolation
of the linear fit shows that the balanced temperature is located near
$0^o C$, i.e.\ well below the temperature of the main transition.

USS gratefully acknowledges support by the Minerva foundation.

\begin{table}
\begin{tabular}{|l|l|l|l|l|l|l|l|} \hline
& G & D & P & I-WP & S & F-RD & C(P) \\ \hline \hline             
$\Delta$ & $0.219$ & $0.219$ & $0.219$ & $0.483$ & $0.587$ & $0.650$ & $0.842$ \\ \hline
$\Gamma$ & $0.767$ & $0.750$ & $0.716$ & $0.743$ & $0.795$ & $0.654$ & $0.656$ \\ \hline
\end{tabular}
\caption{Universal quantities, which characterize the geometry  
of minimal surfaces. $\Delta$ is the width of the distribution 
of Gaussian curvature and determines the overall stability of 
bicontinuous cubic phases. The values for $\Delta$ 
are the same for $G$, $D$ and $P$ due to a
Bonnet transformation. The \emph{topology index} $\Gamma$ describes
the porosity and determines the position in the phase diagram.}
\label{TableNumbers}
\end{table} 

\end{multicols} 

\end{document}